\documentclass[draft]{agujournal2018}
\usepackage{apacite, amsmath}
\usepackage{url} 
\usepackage{lineno}
\draftfalse

\journalname{JGR: Space Physics}

\def\aa{{\it Astron. \& Astrophys. Lett.}}

\newcommand{\aap}{    {\it Astron. Astrophys.}}

\newcommand{\apj}{    {\it Astrophys. J.}}
\newcommand{\apjl}{   {\it Astrophys. J. Lett.}}

\newcommand{\grl}{    {\it Geophys. Res. Lett.}}

\newcommand{\jgr}{    {\it J. Geophys. Res.}}

\newcommand{\nat}{    {\it Nature}}

\newcommand{\solphys}{{\it Solar Phys.}}

\newcommand{\ssr}{    {\it Space Sci. Rev.}}

\begin{document}

%
%


\title{Beyond the mini-solar maximum of solar cycle 24: Declining solar magnetic fields and the response of the terrestrial magnetosphere}

%
%




\authors{M. Ingale\affil{1}, P. Janardhan\affil{1}, and S. K. Bisoi\affil{2}}


\affiliation{1}{Physical Research Laboratory, Astronomy \& Astrophysics
	Division, Navarangpura, Ahmedabad - 380 009, India.}

\affiliation{2}{Key Laboratory of Solar Activity, National Astronomical
	Observatories, Chinese Academy of Sciences, Beijing 100012, China}




\correspondingauthor{M Ingale}{31mayur83@gmail.com}




\begin{keypoints}
\item Global response of the terrestrial magnetosphere to the
long term decline in solar magnetic fields
\item Steady increase of terrestrial magnetopause and bow shock stand-off distance since around mid-1990's
\item Forecast of an expanded terrestrial magnetopause
during the minimum of present solar cycle 24
\end{keypoints}

%
%


\begin{abstract}
The present study examines the response of the terrestrial
magnetosphere to the long-term steady declining trends observed
in solar magnetic fields and solar wind micro-turbulence levels
since mid-1990's that has been continuing beyond the mini-
solar maximum of cycle 24. A detailed analysis of the response
of the terrestrial magnetosphere has been carried out by
studying the extent and shape of the Earth's magnetopause and
bow shock over the past four solar cycles. We estimate sub-solar
stand-off distance of the magnetopause and bow shock, and the
shape of the magnetopause using numerical as well as empirical
models. The computed magnetopause and bow shock stand-off
distances have been found to be increasing steadily since around
mid-1990's, consistent with the steady declining trend seen in solar
magnetic fields and solar wind micro-turbulence levels. Similarly,
we find an expansion in the shape of the magnetopause since 1996.
The implications of the increasing trend seen in the magnetopause
and bow shock stand-off distances are discussed and a forecast
of the shape of the magnetopause in 2020, the minimum of cycle 24,
has been made. Importantly, we also find two instances between
1968 and 1991 when the magnetopause stand-off distance dropped
to values close to 6.6 earth radii, the geostationary orbit, for
duration ranging from 9$-$11 hours and one event in 2005, post
1995 when the decline in photospheric fields began. Though there
have been no such events since 2005, it represents a clear and
present danger to our satellite systems.

\end{abstract}

%
%

%
\section{Introduction}

The long term variability in solar magnetic fields can induce
changes in the terrestrial magnetosphere, with the solar wind
providing the complex link through which the effect is mediated.
This link has been of particular interest to the solar and space
science community due to the peculiar behaviour seen in
solar cycles 23 and 24 and in view of the long term changes taking
place on the sun and in the solar wind over the past three solar
cycles.  The variation of photospheric fields obtained from
ground-based magnetic field measurements since 1970's have been
used as an indicator of solar cycle activity.  Our studies of solar
photospheric magnetic fields \citep{JBG10,BiJ14,JaB15,JaF18}
in the past few years have shown a steady and continuous decline of
solar high-latitude photospheric fields since mid-1990's.  In
earlier studies, we have also shown the steady decline in solar wind
micro-turbulence levels in the inner heliosphere, spanning
heliocentric distances from 0.2 to 0.8 AU \citep{JaB11,JaB15,BiJ14b},
in sync with the decline in photospheric magnetic fields. The
solar wind turbulence levels were obtained using ground based
interplanetary scintillation (IPS) measurements. IPS essentially
provides one with an idea of the large scale structure of the
solar wind \citep{ACK80, ABJ95}.  Early, IPS measurements however,
were primarily employed in determining angular sizes of radio
sources \citep{RHe72,JAl93}. More recent observations have
provided deep insights into the global structure of the solar
wind and heliospheric magnetic field (HMF) all the way out to
the solar wind termination shock at $\sim$90 AU, where 1 AU
is the sun-earth distance \citep{FuT16}. The long term declining
trends seen in both photospheric magnetic fields and solar wind
micro-turbulence levels over the entire inner-heliosphere, coupled
with the unusually deep solar minimum in cycle 23 and the very
unusual solar polar field conditions seen in cycle 24
\citep{GYa16,JaF18}, implies that these changes could directly
affect the terrestrial magnetosphere.

One way to quantify the effect of the long term changes in solar
cycle activity on the terrestrial magnetosphere (MS) is by
investigating the size and shape of the bow shock (BS) and
magnetopause (MP) under different solar wind conditions. The
BS forms in the upstream region of the Earth's magnetosphere,
followed by the magnetosheath which in turn is bounded on the earthward side by the MP. The changing shape and position of
the BS and MP with time is important in understanding space
weather and also in planetary exploration because much like
the earth, other planetary magnetosphere would have also
undergone changes in their MP shape as a result of the observed
global changes occurring in the solar wind.
Using empirical and numerical models, the location
and shape of the MP and BS under various solar wind conditions
have been predicted since a long time.  Most of the empirical
models \citep{Fai71, For79, NSa91, SLR91, ShC97, ShS98,
BoE00, LiZ10, JNS12}, with a few exceptions \citep{WaS13, SGo15},
assume a priori functional form for the MP and then estimate
the free parameters using available spacecraft crossing database.
For example, \citet{ShS98} proposed the following functional
form,
\begin{equation}
r = r_{mp} \left(\frac{2}{1 + \cos\theta}\right)^{\alpha} \,\,\, R_E.
\label{mp_position}
\end{equation}
Equation (\ref{mp_position}) has two parameters ${\it{viz.}}$
r${_{mp}}$, the MP stand-off distance and $\alpha$, the flaring
parameter. Angle $\theta$ is the solar zenith angle (the angle
between the Sun-Earth line and the radial direction) of the
point of interest. This assumption can, however, introduce
errors in the cusp regions of Earth's magnetosphere due to
the asymmetric MP shape at that location.  On the other hand, most
of the numerical models used to determine the
MP position and shape \citep{CLy95, PeS95, EWi97, CCa03, GHu07},
were based on non-axis symmetric global MHD simulations.  So, unlike
empirical models, the condition of pressure balance between the
solar wind dynamic pressure and the pressure due to the earth's
dipole magnetic field \citep{CFe31, ZRo59, Bea60, SBr62, MBe64, Ols69}
is satisfied at every point.  However, the numerical models
generally don't include all magnetospheric current systems and
often use an impermeable, infinitely conducting MP as an obstacle,
which is far from reality. The BS shape and location is usually
determined by the shape and location of MP as well as by various
solar wind conditions, like for example the solar wind ram
pressure, the magnetosonic and Alfv\'en Mach numbers and the
orientation of the interplanetary magnetic field (IMF). Several empirical \citep{Fai71,For79,FRu94,PeS95,FaC01,JNS12} and
numerical models \citep{CLy95,CLy96,CGr94,CCa03} have been used
to estimate the shape and location of the BS.
%
\begin{figure}
	\centering
	\vspace{0.4cm}
	\includegraphics[width=0.8\textwidth]{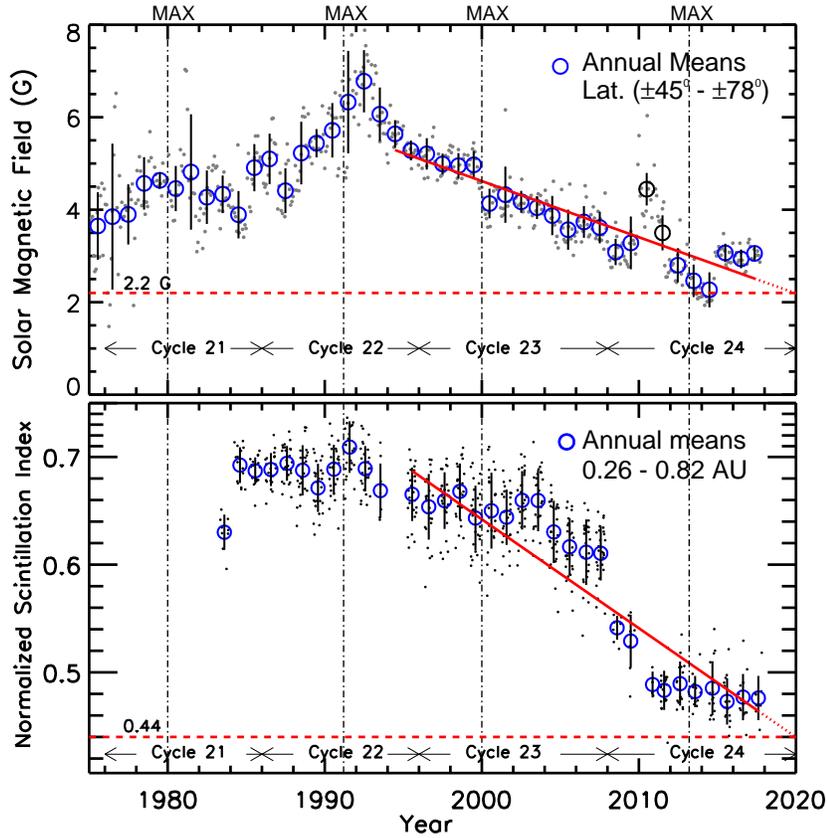}
	\caption{Solar high-latitude photospheric magnetic fields in
		the latitude range $45^{\circ}--78^{\circ}$ for the period
		Feb. 1975$-$Dec. 2017 (upper panel) and solar wind
		micro-turbulence levels from 327 MHz IPS observations
		in the period 1983$-$2017 and in the distance range
		0.2$-$0.8 AU (lower panel). The filled grey dots in
		both panels are actual measurements of magnetic fields
		(top) and solar wind micro-turbulence (bottom), while
		the open blue circles are annual means shown with
		1$\sigma$ error bars. The solid red line in both
		panels is a best fit to the declining trends for the
		annual means while the dotted red lines are
		extrapolations of the best fit until 2020 for the
		photospheric fields (top) and the IPS observations (bottom).
		The horizontal red dashed lines are marked at 2.2 G and 0.44
		in the upper and lower panels, respectively.}
	\label{fig1}
\end{figure}

The actual shape or position for the BS and MP is however still far
from reality. We have therefore used both empirical and numerical
models together to study the position and shape of MP and BS by
determining the stand-off distances of MP and BS as a function of
time over the last four decades, $1975 - 2017$, and have
in turn linked them to the long-term trends in solar magnetic activity. It is to be noted however, that our study concentrates
only on estimating the long-term trends in the MP and BS stand-off distances and the shape of the MP. Therefore, in this paper, we estimate the MP stand-off distance using models by \citet{LiZ10},
and \citet{LuL11}, as being representative of empirical and
numerical models, respectively see the model description in \S 3.1). Similarly, the bow shock and it's stand-off distance has been
estimated using \citet{JNS12} and \citet{CCa03} as being representative of empirical and numerical models, respectively
see the model description in \S 3.2). The other study that we
found for such long-term trends for the MP stand-off distances
is by \citet{McA13}, wherein, the authors reported the canonical stand-off distance of the MP to be about 11 earth radii (R$_{E}$),
for the period 2009 to 2013, covering the minimum of cycle 23
to the early rise phase of cycle 24, compared to about 10 R${_{E}}$ for the period 1974 to 1994, covering cycles 21--22. While the
normal value of the MP is usually $\sim$10 R$_{E}$. Hence, it is evident that the long term study of the shape and location of the
BS and MP is important. Recently, \citet{SaB19}, using solar wind observations and empirical magnetopause models, also reported an increase in average annual magnetopause stand-off distance by
nearly 2 R$_{E}$ between 1991 (9.7 R$_{E}$) and 2009 (11.6 R$_{E}$).
In this study we therefore, attempt to quantify the changes in
shape and location of the Earth's magnetosphere over the last
four solar cycles, and to compare the results obtained using both
empirical and numerical models.
\section{Observations}
\subsection{Solar photospheric magnetic fields}
Figure \ref{fig1} (upper panel) shows solar photospheric magnetic
fields in the latitude range $45^{\circ}$-$78^{\circ}$ for the period
Feb.1975$-$Dec.2017. Here, we selected the latitude range
$45^{\circ}$-$78^{\circ}$ to represent the high-latitude solar
polar fields. It must however, be kept in mind that different
researchers have used different latitude ranges to estimate
polar fields ${\textit{viz.}}$ poleward of $45^{\circ}$ \citep{JaB11,BiJ14}, $55^{\circ}$ (Wilcox Solar Observatory
polar fields, \url{http://wso.stanford.edu/Polar.html},
\cite{JaF18}), $60^{\circ}$ \citep{Tom11,GoY12,SuH15,GYa16},
and $70^{\circ}$ \citep{MuS12}. The other point to note here
that the polar fields studied by different researchers are
actually the signed values of solar polar fields (see Figure
1 in \cite{JaF18}), which reverse polarity around solar maximum
of each solar cycle. However, in the present study, we emphasize
only on the unsigned or absolute values of solar polar fields
in the latitude range $45^{\circ}$-$78^{\circ}$. It is seen
(see Figure 3 in \cite{JaB15} and Figure 5 in \cite{JaF18})
that the unsigned values of solar magnetic fields in the latitude
range $0^{\circ}$-$45^{\circ}$ follow the solar cycle with the
fields attaining their maximum at around solar maximum of each
solar cycle.

The filled grey dots in the upper panel of Fig.\ref{fig1}
are unsigned or absolute values of magnetic field
measurements for each Carrington-rotation (CR) while the
open blue circles are annual means, of Carrington rotation
averaged magnetic fields, shown with 1$\sigma$ error
bars. The unsigned Carrington-rotation averaged solar magnetic
fields were computed using synoptic magnetograms from
CR1625--CR2197, available as standard FITS format files,
obtained from the National Solar Observatory, Kitt Peak (NSO/KP)
and the Synoptic Optical Long-term Investigations of the Sun (NSO/SOLIS) facility. Each synoptic magnetograms represents
arrays of 360 $\times$ 180 in longitude and sine of latitude
format with magnetic field distribution for each Carrington
rotation or 27.2753 day averaged photospheric magnetic fields
in units of Gauss. Details about the computation of magnetic
fields can be referred to in \cite{JBG10,JaF18}.

It is evident from Fig.\ref{fig1} (upper panel) that the
decline in the field strength of high latitude photospheric
magnetic fields that has begun since $\sim$1995 has been
continuing till the end of Dec. 2017. However, for the years
2010-2011, a sudden increase in the field strength is first
observed, while a similar increase is then observed for the
years 2015-2017.  It is to be noted however, that the
decreasing trend in the field strength still persists.
The 25-year long decreasing trend seen in the
high-latitude field strength shows an $\sim$40\% drop from its peak
value in 1995 to the value in 2017. In Fig.\ref{fig1} (upper panel),
the solid red line is a best fit, with a Pearson correlation
coefficient of $r = -0.91$ at a significance level of $99\%$, to
the declining trend of high-latitude field strength for all the
annual means in the period 1995--2017.  Based on magnetic field
measurements from Feb. 1975\,--\,Jul. 2014, \cite{JaB15} reported
that the trend of declining high-latitude field strength would
continue in the same manner up to the minimum of the current solar
cycle 24, ${\it{i.e.}}$ $\sim$2020. In fact, as evident from
Fig.\ref{fig1}, it is seen that the declining trend in
the high latitude fields have been continuing. Thus,
we expect the trend to continue in the same manner and extrapolate
the solid red line beyond 2017 upto 2020, the expected minimum of
the current solar cycle 24 as shown by the dotted red
line, in Fig.\ref{fig1}. The expected value of high-latitude
field strength in 2020 is $\sim$2.2 ($\pm$0.08) G, as indicated
in the upper panel of Fig. \ref{fig1} by a dashed red horizontal
line. A clear long term decreasing trend in solar high-latitude
field strength is apparent and it is reasonable to assume that
this trend will continue at least until 2020, the expected
minimum of the current cycle.
\subsection{Solar wind micro-turbulence levels}

The lower panel in Fig.\ref{fig1} shows the solar wind
micro-turbulence levels as measured by 327 MHz IPS observations
for the period 1983 to the end of 2017 and in the heliocentric
distance range 0.2$-$0.8 AU, covering the inner heliosphere.
These measurements were made using  the three station IPS
observatory of the Institute for Space-Earth Environmental
research (ISEE), Nagoya, Japan. The measurements of scintillation
index obtained from ISEE usually vary with the heliocentric
distance and with the angular diameter of the observed radio
source. As a result, the scintillation index measurements for
a selected number of compact, point-like extra-galactic radio
sources in the period 1983--2017 were normalized in a manner
to make them independent of heliocentric distance and angular
source size such that they should show a scintillation index
of unity at a radial distance of 0.2 AU (for more details see
\cite{JaB11} and \cite{BiJ14b}). The  filled grey dots in the
lower panel of Fig. \ref{fig1} are measurements of normalized scintillation index while the open blue circles are annual
means shown with 1$\sigma$ error bars. The solid red line is
a least square fit to the declining trend for the annual means
with a Pearson correlation coefficient of $r = -0.93$, at a
significance level of $99\%$.

It is seen from Fig.\ref{fig1} (lower panel) that the
normalized scintillation level has been steadily declining, in
sync with the declining photospheric fields, until
the end of 2017. The implication of the declining scintillation
index is that a strongly scintillating point-like, extra-
galactic radio source at 327 MHz would appear to scintillate
like a much weaker and extended source. This is due to the
significant decrease in the rms electron density fluctuations
$\Delta$N in the solar wind over time. The fit to the normalized
scintillation index values has been extrapolated beyond 2017 and
is shown by a dotted red line in the lower panel of Fig.\ref{fig1}.
If the decline continues, the normalized scintillation index will
reach a value of 0.44 by 2020, indicated by a horizontal red line
in the lower panel of Fig.\ref{fig1}. This is equivalent to IPS
observations of an extra-galactic source with an angular diameter
of $\sim$190 milliseconds (mas). So a point source like
1148-001, with a measured angular size, using VLBI observations,
of 10 mas will scintillate like a 190 mas source by 2020.
\subsection{Solar wind parameters at 1 AU}
\label{swp}
%
%
\begin{figure}
	\centering
	\vspace{0.4cm}
	\includegraphics[width=0.4\textheight]{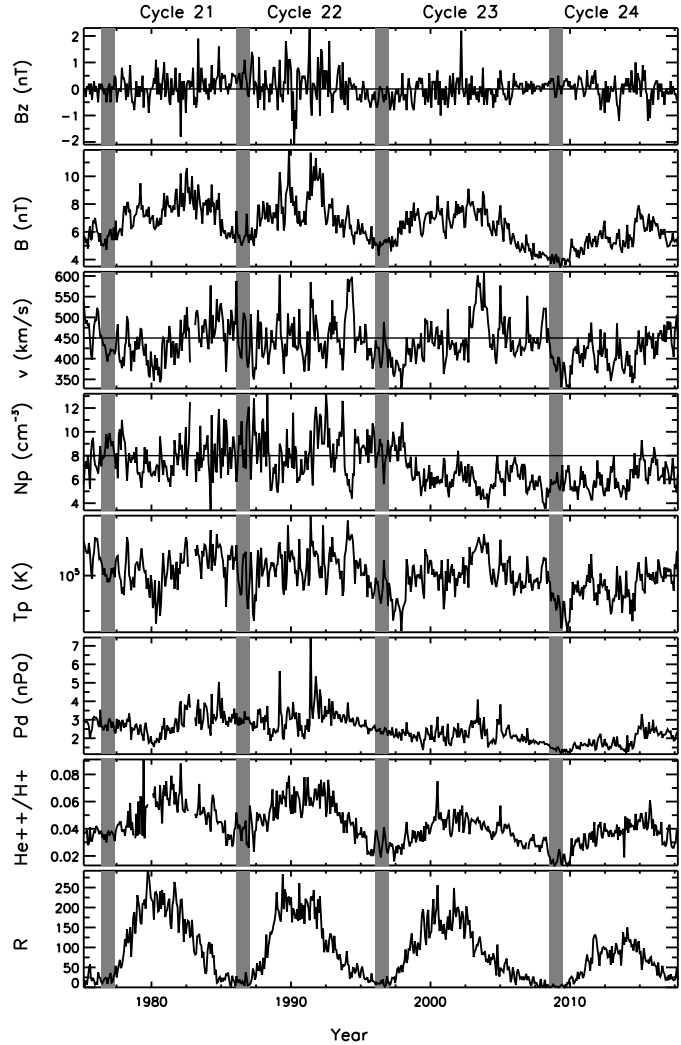}
	\caption{Carrington rotation-averaged insitu measurements
		of solar wind parameters at 1 AU (first to eighth panels)
		B${_{z}}$, B, v, N${_{p}}$, T${_{p}}$, P${_{d}}$, He$^{++}$/H$^{+}$, and R, respectively, for the period Feb. 1975--Dec. 2017, covering solar cycles 21--24. The vertical gray bars demarcate 1 year intervals during the minima of solar cycles 20--23. The solid horizontal lines in third
		and fourth panels are drawn to indicate the average values
		of v and N${_{p}}$, respectively.}
	\label{fig2}
\end{figure}
%
The eight panels in Figure \ref{fig2} show Carrington-rotation
(or 27.2753 day) averaged solar wind parameters at 1 AU.
Starting from the top and going down, the eight panels
show respectively, the north-south component of IMF (B${_{z}}$)
in GSM coordinate system, the strength of IMF (B), flow speed (v),
proton density (N${_{p}}$), proton temperature (T${_{p}}$), dynamic
pressure(P${_{d}}$) and alpha-to-proton number density ratio (He$^{++}$/H$^{+}$) and the Carrington-rotation averaged
sunspot number(R). The data shown span the period Feb. 1975$-$Dec. 2017, covering solar cycles 21--24.  The gray vertical bands in Fig.\ref{fig2} indicate 1 year intervals during solar minima of
cycles 20--23 corresponding to CR1642–1654, CR1771--1783, CR1905--1917, and CR2072--2084, respectively \citep{JaB15}.
The Carrington-rotation averaged solar wind parameters at 1 AU
presented in Fig.\ref{fig2} were deduced from daily measurements
obtained from the OMNI data base \url{http://gsfc.nasa.gov/omniweb}
while the Carrington-rotation averaged sunspot number were derived
from daily measurements of sunspot number obtained from the Royal
Observatory Belgium, Brussels \url{http://www.sidc.be/silso/datafiles}. The Carrington-rotation
averaged values were derived in order to compare them with the
Carrington-rotation averaged photospheric magnetic fields
as shown in Fig.\ref{fig1} (upper panel). The values of B,
P${_{d}}$ and He$^{++}$/H$^{+}$ in Fig.\ref{fig2} show a 11-year
temporal variation. However, the temporal variation is not a
one-to-one correspondence with sunspot activity variation. The
important point to note from temporal variations of B,
N${_{p}}$, v, P${_{d}}$ and He$^{++}$/H$^{+}$ over the
last four cycles is a long term decreasing trend that has
begun around the mid-1990's and has been continuing till date.
The value of B shows a decrease from $\sim7.8$ nT in 1995 to
$\sim 5.2$ nT in 2017, with an average decline of $\sim30\%$.
The solar wind speed is also very low during cycle 24 with
speeds being less than $\sim$450 km/s, most of the time as
indicated by a horizontal line in the third panel of Fig.\ref{fig2}.
Not only the solar wind speed, but also we also see a
significant reduction in solar wind proton density, especially
starting from cycle 23 and continuing in current cycle 24 until
the present. For most of the time, the value of proton density
remained below $\sim$8 cm$^{-3}$, indicated by a horizontal
line in the fourth panel of Fig.\ref{fig2}. While the average
decline in P${_{d}}$, over the last $\sim$20 years,
was found to be $\sim$40\%. \cite{McA13} reported the weakest
solar wind conditions during the maximum of current cycle 24
(their data spanned the period up to the maximum of
cycle 24) and referred to the cycle 24 maximum as
mini-solar maximum. Although, as seen from SSN in the eight
panel of Fig.\ref{fig2}, we are approaching the minimum of cycle
24, however, the weak nature of the solar wind as a whole has
been still continuing. In addition, it is to be noted that the
values of all solar wind parameters have remained at an all time
low during the extended solar minimum of cycle 23. The long term
decreasing trend of solar wind parameters at 1 AU is in
correspondence with the diminishing sunspot cycle activity
seen since cycle 22. The changes in solar wind conditions at
1 AU observed in the recent period is in keeping with the
declining photospheric magnetic fields and solar wind
micro-turbulence levels in the inner heliosphere that began
around the same period.
%
\begin{center}
	\begin{figure}[ht]
		\centering
		\vspace{0.6cm}
		\includegraphics[width=0.4\textheight]{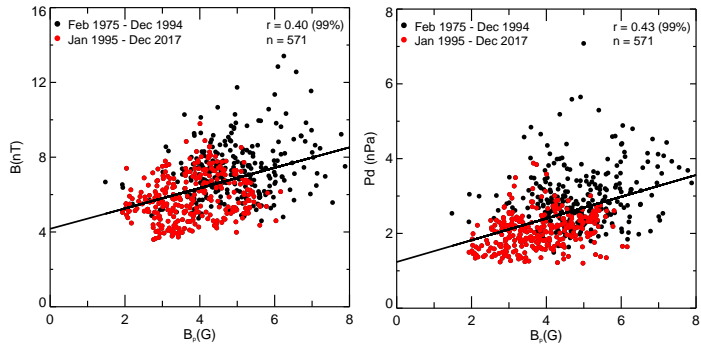}
		\caption{B (left panel) and P${_{d}}$ (right panel) as
			function of B${_{p}}$ for the period Feb. 1975--Dec. 2017.
			The respective Pearson's correlation coefficients
			of r = 0.40 and 0.43 with confidence intervals of 99\%
			are indicated in the top right corners of the panels.
			The filled black and red dots are measurements for
			the period Feb. 1975--Dec.1994 and Jan. 1995--Dec.
			2017, respectively while the solid black lines in
			the panels are best fits to the all of the data
			points (both black and red dots).}
		\label{fig3}
	\end{figure}
\end{center}
%
In order to show a causal link of B and P${_{d}}$ with solar
high-latitude fields (B${_{p}}$), we plotted the correlation
of B with B${_{p}}$ in the left panel of Figure \ref{fig3} and
that of P${_{d}}$ with B${_{p}}$ in the right panel of
Fig.\ref{fig3}. The solid black and red dots shown in Fig.\ref{fig3}
are, respectively, the Carrington-rotation (27.2753 days)
averaged data points for the period 1975\,--\,Dec. 1994 and
Jan. 1995\,--\,Dec. 2017. It is to be remembered here
that B${_{p}}$ have shown a steady decline since around 1995.
The data points are, therefore, shown in solid black and red
dots so as to compare the correlation of B and P${_{d}}$ with
B${_{p}}$, prior to 1995 and after 1995.
It can be seen from Fig.\ref{fig3} (left panel) that B show a
correlation with B${_{p}}$ with a Pearson correlation
coefficients of $r = 0.40$ at a significance level of $99\%$,
indicated at the top right corner of the left panel of
Fig.\ref{fig3}. Similarly, a correlation between P${_{d}}$ and
B${_{p}}$ with a correlation coefficient of $r = 0.43$ at a
significance level of $99\%$ can be seen from Fig.\ref{fig3}
(right panel).
It is seen, as shown by the red dots, that post-1995 the values
of B and P${_{d}}$ have decreased significantly. The solid black
line in each panel in Fig.\ref{fig3} is a best fit to all data
points (both black and red dots) of B and P${_{d}}$ with
B${_{p}}$. The significant decrease in values of B and P${_{d}}$
post mid-1990's can be attributed to the weaker solar wind
conditions in the inner heliosphere. The signatures of this
weakening have already been shown in the solar wind turbulence
levels in the inner heliosphere, as shown in Fig.\ref{fig1}
(lower panel). \cite{JaB15} reported a good correlation between
B and B${_{p}}$ at solar minima using measurements for cycles
20--23. Here, we have revisited the correlation for B and
B${_{p}}$ at solar minima, using Carrington-rotation averaged
data for 1 year intervals at each solar minimum of
cycles 20--23, corresponding to the Carrington-rotation
period of CR1642--1654, CR1771--1783, CR1905--1917, and
CR2072--2084, respectively. The correlation of B and B${_{p}}$
during solar cycle minima is plotted in the left panel
of Fig.\ref{fig4}. Similarly, the correlation of P${_{d}}$
with B${_{p}}$ during solar cycle minima is plotted in the
right panel of Fig.\ref{fig4}. The respective Person's
correlation coefficients, r = 0.50 and 0.62, are indicated
in the top right corner of each panel. It is evident from
the correlation
%
\begin{center}
	\begin{figure}[ht]
		\centering
		\vspace{0.6cm}
		\includegraphics[width=0.4\textheight]{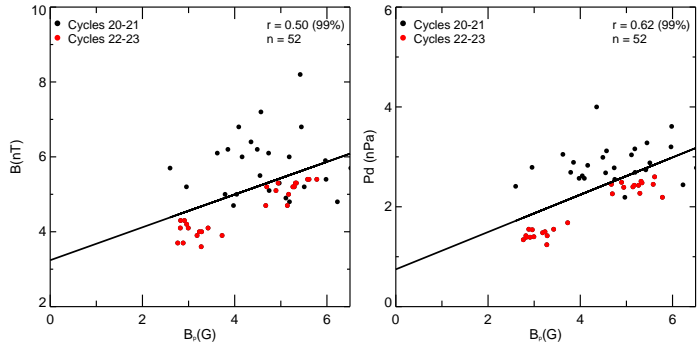}
		\caption{B (left panel) and P${_{d}}$ (right panel)
			as function of B${_{p}}$ for one year interval period
			during the minima of cycles 20--23. The respective Pearson’s correlation coefficients, r = 0.50 and 0.62, with confidence intervals are indicated in the top
			right corner of each panel, respectively. The filled
			black and red dots in each panel are measurements
			for cycles 20--21 and cycles 22--23, respectively.
			The solid black lines are best fit to all of the
			data points (both black and red dots).}
		\label{fig4}
	\end{figure}
\end{center}
%
coefficient values that B and P${_{d}}$ show a good correlation
with B${_{p}}$. While in order to show the global reduction of
B and P${_{d}}$ post mid-1990's, we plotted the data points
for cycles 20--21 and cycles 22--23 using the filled black and red
dots, respectively. It can be seen that post mid-1990's, i.e., for
cycles 22 and 23, as shown by the red dots in each panel,
we see a significant decrease in the strength of B and P${_{d}}$.
Also, we found better correlations for B and P${_{d}}$ with B${_{p}}$
for data points covering the minima of cycles 22 and 23
with correlation coefficients of r = 0.92 and 0.95, respectively.
From Figures \ref{fig2}, \ref{fig3} and \ref{fig4}, it is clear
that the values of B and P${_{d}}$ have shown significant reductions
during cycles 22--24, i.e., since around mid-1990's when solar
high-latitude fields started to decline. As mentioned earlier,
the location (or subsolar stand-off distance) of BS and MP depend
on strength of P${_{d}}$ which show good correlations with solar
polar fields, it is therefore important to know how
the subsolar stand-off distances of BS and MP are related to the
long term declining trend seen in solar polar fields
since mid-1990's. In the following sections, we discuss the
estimation of stand-off distances of BS and MP, and their temporal
variation in keeping with the changes in solar polar fields.
\section{Methodology} \label{sec:models}
\begin{center}
	\begin{figure}[ht]
		\centering
		\includegraphics[width=0.4\textheight]{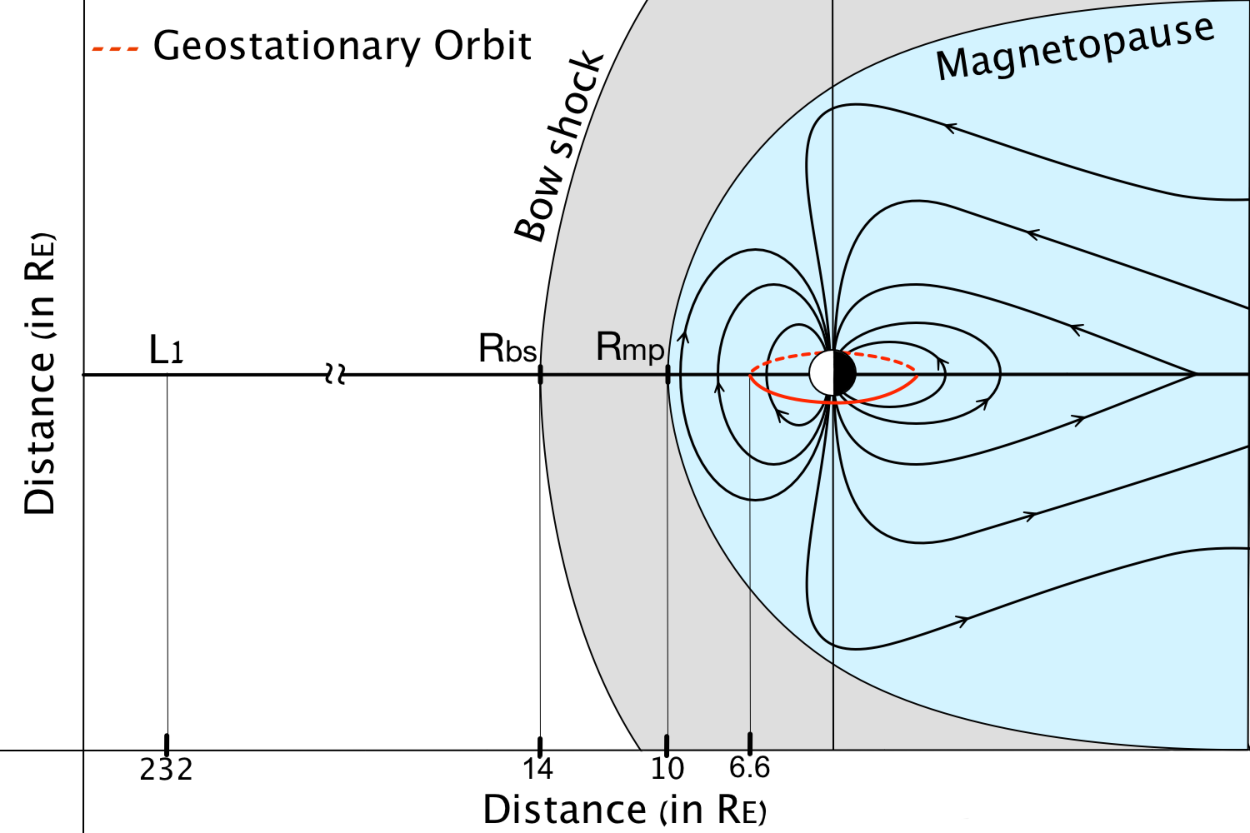}
		\caption{A schematic of the stand-off distance of the
			MP ($R_{\textrm{mp}}$) and BS ($R_{\textrm{bs}}$)
			in the GSM coordinate system.  The dotted red circle
			in the equatorial plane represents the geostationary
			orbit at 6.6 earth radii. The L1 Lagrangian point of
			the sun-earth system is at 232 earth radii.}
		\label{fig5}
	\end{figure}
\end{center}

Figure \ref{fig5} shows a schematic representation (not to scale)
of the position and shape of the MP and BS in the GSM
coordinate system wherein, the earth is considered to be at the
origin. The x-axis is along the sun-earth line, the z-axis
is perpendicular to the plane of the earth's orbit. $R_{\textrm{bs}}$
and $R_{\textrm{mp}}$ represent stand-off distances of the BS and MP,
respectively. The nominal positions of the stand-off distances of
the MP and BS at 10 and 14 earth radii (R${_{E}}$) respectively,
are indicated. Also shown, by a red circle at 6.6 R${_{E}}$, is
the geostationary orbit in the earth's equatorial plane and the
location of the Lagrangian point, L1, of the sun-earth system at
232 R${_{E}}$.

\subsection{MP stand-off distance}

We briefly describe below the models used in the
calculations of the MP stand-off distance and MP shape.
The first model is by \cite{LiZ10} which represents an
empirical approach, whereas the second model by \cite{LuL11} is
representative of a numerical approach.
%
%
\subsubsection{Empirical model}

\cite{LiZ10}, abbreviated as L10 hereafter, extended the
assumptions of \cite{ShS98} to address asymmetries and
indentations near the polar cusps. Employing a database
of nearly 2708 MP crossings from observations by Cluster,
Geotail, Goes, IMP8, Interball, LANL, Polar, TC1, THEMIS,
WIND and Hawkeye, along with corresponding solar wind
parameters from ACE, Wind and OMNI, they obtained a model
for the MP which was parametrized by the solar wind dynamic
and magnetic pressure (P${_{d}}$ + P${_{m}}$), IMF-B${_{z}}$
and dipole tilt angle ($\psi$), the dipole magnetic
latitude of the subsolar point. Based on the relation between
P${_{d}}$ and r${_{mp}}$, the influence of IMF B${_{z}}$ on
r${_{mp}}$ \citep{ShS98} and the saturation effect of a
southward IMF B${_{z}}$ on r${_{mp}}$ \citep{YaC03}, L10
expressed the stand-off distance for MP as:
\begin{equation}
r_{mp} = a_0(P_d + P_m)^{a_1}\left(1 + a_2 \frac{\textrm{exp}
	(a_3 B_z)-1}{\textrm{exp}(a_4 B_z) + 1}\right) \,\,\, R_E
\label{L10}
\end{equation}
The coefficients ($a_0$, $a_1$, $a_2$, $a_3$ and $a_4$) in
equation (\ref{L10}) are obtained by using non-linear multi
parameter fitting (Levenberg $-$ Marquardt method) based on
observations of 247 MP crossings that were found near the
stand-off distance. The coefficients are listed in Table$-$2
of L10.

To obtain the MP shape, L10 expanded the eq. (\ref{mp_position})
as,
\begin{equation}
r = r_{mp} \left \{\mathrm{cos} \frac{\theta}{2} + a_5 \cdot
\mathrm{sin}(2 \theta) [1 - \mathrm{exp}(-\theta)] \right \}
^{\beta} \,\, R_E.
\label{mp_shape_L10}
\end{equation}
where the factor $1-exp(-\theta)$ smooths out the MP shape near
the subsolar point.  The asymmetries and indentations are introduced
through the azimuthal angle $\phi$, the angle between the projection
of r in Y-Z plane and the direction of positive Y axis.  The flaring
parameter $\beta$ (given by equation (5) of L10) is,
\begin{equation}
\beta = \beta_0 + \beta_1 \mathrm{cos}(\phi) + \beta_2 \mathrm{sin}(\phi)
+ \beta_3 \mathrm{sin}^2(\phi).
\label{beta}
\end{equation}
We considered the simpler case: $\phi=0$ (meridional plane)
for which equation (\ref{beta}) reduces to $\beta = \beta_0 +
\beta_1$. $\beta_0$ and $\beta_1$ (eq. 16 and 17 of L10) are
obtained using observations of 422 MP crossings. The relevant
values of the parameters are listed in Table$-$6 of L10.

The L10 model (eqs \ref{L10} and \ref{mp_shape_L10}) yields good
results in predicting the MP stand-off distance and shape. Also,
when compared with the observed low latitude MP crossings, the
standard deviation of the L10 model is the least (0.54 $R_E$)
amongst several other models \citep{JNS12}.  We, therefore,
prefer \cite{LiZ10} model for the calculations of MP stand-off
distance and shape.

\subsubsection{Numerical model}
To estimate the MP stand-off distance and shape using
numerical calculations, we preferred the model by
\cite{LuL11}, abbreviated as L11 hereafter, based on global
MHD simulations that use the Space Weather Modelling Framework
(SWMF), a framework for physics-based space weather
simulations \citep{ToS05}.  The functional form of \cite{ShS98}
was extended to describe the global MP size and shape using the
method of multi-parameter fitting. L11 included
azimuthal asymmetry via $\phi$ and extended the functional
form shown in eq. (\ref{mp_position}). Thus, the dayside MP
is given by,
\begin{equation}
r = r_{mp} \left(\frac{2}{1 + \mathrm{cos}\theta} \right)
^{\alpha + \beta_1 \mathrm{cos}\phi} \,\, R_E.
\label{mp_shape_L11}
\end{equation}
where $\beta_1$ characterizes the azimuthal asymmetry with
respect to $\phi$. Using the fitting results from \cite{ShC97},
the relationship between the (r${_{mp}}$, $\alpha$, $\beta_1$)
and solar wind conditions (P${_{d}}$, B${_{z}}$) were evaluated.
The multiple parameter fitting results in the following best-fit
functions (eq., 18, 19, 20 of L11):
\begin{equation}
r_{mp} =
\begin{cases}
(11.494 + 0.0371 B_z) P_d^{-1/5.2}, & B_z \geq 0 \\
(11.494 + 0.0983 B_z) P_d^{-1/5.2}, & B_z < 0
\end{cases}
\label{r0_L11}
\end{equation}
\begin{equation}
\alpha =
\begin{cases}
(0.543 - 0.0225 B_z + 0.00528 P_d + 0.00261 B_z P_d),
{\mathrm{for}} \, B_z \geq 0 \\
(0.543 + 0.0079 B_z + 0.00528 P_d - 0.00019 B_z P_d),
{\mathrm{for}} \, B_z < 0
\end{cases}
\label{alpha_L11}
\end{equation}
\begin{equation}
\beta_1 =
\begin{cases}
(-0.263 + 0.0045 B_z - 0.00924 P_d - 0.00059 B_z P_d),
{\mathrm{for}} \, B_z \geq 0 \\
(-0.263 - 0.0259 B_z - 0.00924 P_d + 0.00256 B_z P_d),
{\mathrm{for}} \, B_z < 0
\end{cases}
\label{beta_L11}
\end{equation}
The model results by L11 yield good matching when compared
with the high and low latitude MP crossings. So we prefer
to use the numerical model by L11.

\subsection{BS stand-off distance} \label{subsec:BS}

We briefly describe below the models used in the
calculations of the BS stand-off distance. The first model,
\cite{JNS12} represents an empirical approach while the second
one, \cite{CCa03}, represents a numerical approach.
\subsubsection{Empirical model}
The model by \cite{JNS12}, hereafter abbreviated as J12,
investigated the solar wind, magnetosheath and magnetosphere
using measurements from the {\textit{THEMIS}} and {\textit{ACE}}
spacecrafts. J12 assumed a parabolic shape for the BS, which is
rotationally symmetric around the $X_{GSM}$ coordinate system.
The authors used an analytic expression, given by
	\begin{equation}
	r_{bs} = 15.02 P_d^{-1/6.55} \,\,\, R_E.
	\label{JRbs}
	\end{equation}
to estimate the BS stand-off distance, where the free
parameters are generally determined using a least
square fitting to the full data set. The equation (\ref{JRbs})
does not take into account the Mach number, however, these simple
model results are found to be in good agreement when compared with
more than 6000 spacecraft crossings of the BS region. We therefore used
the model by \cite{JNS12} to compute the BS stand-off distance.
\subsubsection{Numerical model}
The model by \cite{CCa03}, hereafter abbreviated as CC03,
uses three dimensional MHD simulations from \cite{CLy95}. The
model uses parameters such as P${_{d}}$, the Alfv\'en Mach
number (M${_{A}}$) and the orientation of the IMF ($\theta_{IMF}$)
with respect to the solar wind velocity ($v_{sw}$) direction.
\cite{CCa03} mainly considered two special cases of $\theta_{IMF}
= 45^{\circ}$ and $90^{\circ}$, of which we use $\theta_{IMF} =
90^{\circ}$ to estimate the BS stand-off distance given by
\begin{equation}
r_{bs} = \left(\alpha_0 + \frac{\alpha_1}{M_A} \right)
\left(\frac{P_d}{1.87} \right)^{-1/6} \,\, \textrm{$R_E$}.
\label{CC04}
\end{equation}
Here ($\alpha_0, \alpha_1$) are the fitting parameters obtained by
a least square fitting to the simulated BS locations. The
CC03 model results have been compared with the available spacecraft
crossings close to the nose region of the BS and it is found that for
the near-Earth regime ($-20 R_E < x < 35 R_E$), the model predicted
the BS location very well.

\section{Data and analysis} \label{sec:dat}
For the present study, in order to compute the stand-off
distances of the MP and BS, we use daily averaged measurements
of solar wind dynamic pressure and IMF into equations \ref{L10},
\ref{r0_L11}, \ref{JRbs}, and \ref{CC04} described in \S 3.1
and \S 3.2. The solar wind dynamic pressure was derived
using solar wind proton density ($N_p$) and solar wind velocity
($v_{\textrm{sw}}$). The data for $N_p$, $v_{\textrm{sw}}$, and IMF for the period Feb. 1975$-$Dec. 2017 were obtained from the OMNI
data base (\url{https://omniweb.gsfc.nasa.gov/}). OMNI data are
available as low resolution and high resolution OMNI data sets.
For the high resolution data (1-min and 5-min averages),
interpolations are usually performed on the phase front normal
directions for gap intervals of less than 3 hours, and for the
time shift for gap intervals of less than one hour. So for
this study, we used the low resolution (available as hourly,
daily and 27-day averages) data for which no interpolation
was performed. Actually, the low resolution OMNI data base is
a compilation of hourly averaged near-earth solar wind magnetic
field data and various other solar wind plasma parameters
from several spacecrafts at geocentric or L1 orbit which have
been extensively cross calibrated, and, for some spacecraft
and parameters, cross-normalized
(\url{https://omniweb.gsfc.nasa.gov/html/ow_data.html}). The
available daily average OMNI data are simple average
of these hourly averaged values. We directly used in our
calculations the available daily averaged measurements of IMF
and solar wind plasma parameters from OMNI data base by excluding
the days with missing values and days that have less than 8
hourly average values of IMF and solar wind plasma parameters.
In the next step, from the computed daily averaged values
of MP stand-off distances, we obtained the Carrington rotation
(27.2753 days) and annual averaged values of MP stand-off
distances. In addition, we also used the monthly averages of IMF--${B_{z}}$, P${_{d}}$ and MP stand-off distances to make
a time series forecast using a Auto Regressive Integrated Moving Average (ARIMA) method (see section \S 5.2 and Appendix A).

The stand-off distances of the MP and BS are sensitive to the variations in P${_{d}}$ and IMF--B${_{z}}$. They are also
affected by the Alfv\'en and magnetosonic Mach numbers. The
models that we used in this study mostly consider these
parameters while estimating the stand-off distances of MP and
BS. The angle between the magnetic field and the solar wind
velocity vector is critical in determining the shock location
of the BS \citep{CLy95}.  As mentioned earlier, the shape of
the MP is asymmetric due to the cusp regions. We, therefore,
used models with an analytical formula, taking into account
the pressure imbalance due to the Earth's dipole tilt angle
($\psi$) that reproduce the cusp regions resulting in an
asymmetric MP shape \citep{LiZ10, SGo15}.
\cite{ZhW14} have shown that the earth's dipole moment
has been decaying over the past 1.5 centuries. Even assuming a
linear rate of decay to persist their results suggest that the
average stand-off distance of the MP would only move
$\sim 0.3 R_E$ towards the earth per century.  Since in the
present study, we are looking for the long term trend in the
change of MP and BS stand-off distances, so the contribution
to the MP position and shape at the subsolar point due to the
change in the pressure caused by the declining dipole moment
will be small. Thus, we neglected the effect of the dipole
tilt angle and for the present study, we consider the MP and
BS to be symmetric about the sun-earth line, {\textit{i.e.}}
around the X-axis of the GSM coordinate system.  Further, for
calculating the BS stand-off distance using the numerical model
by CC03 we kept the IMF angle ($\theta$) fixed at $90^{\circ}$.
%
%
\begin{center}
	\begin{figure}[ht]
		\centering
		\includegraphics[width=0.4\textheight]{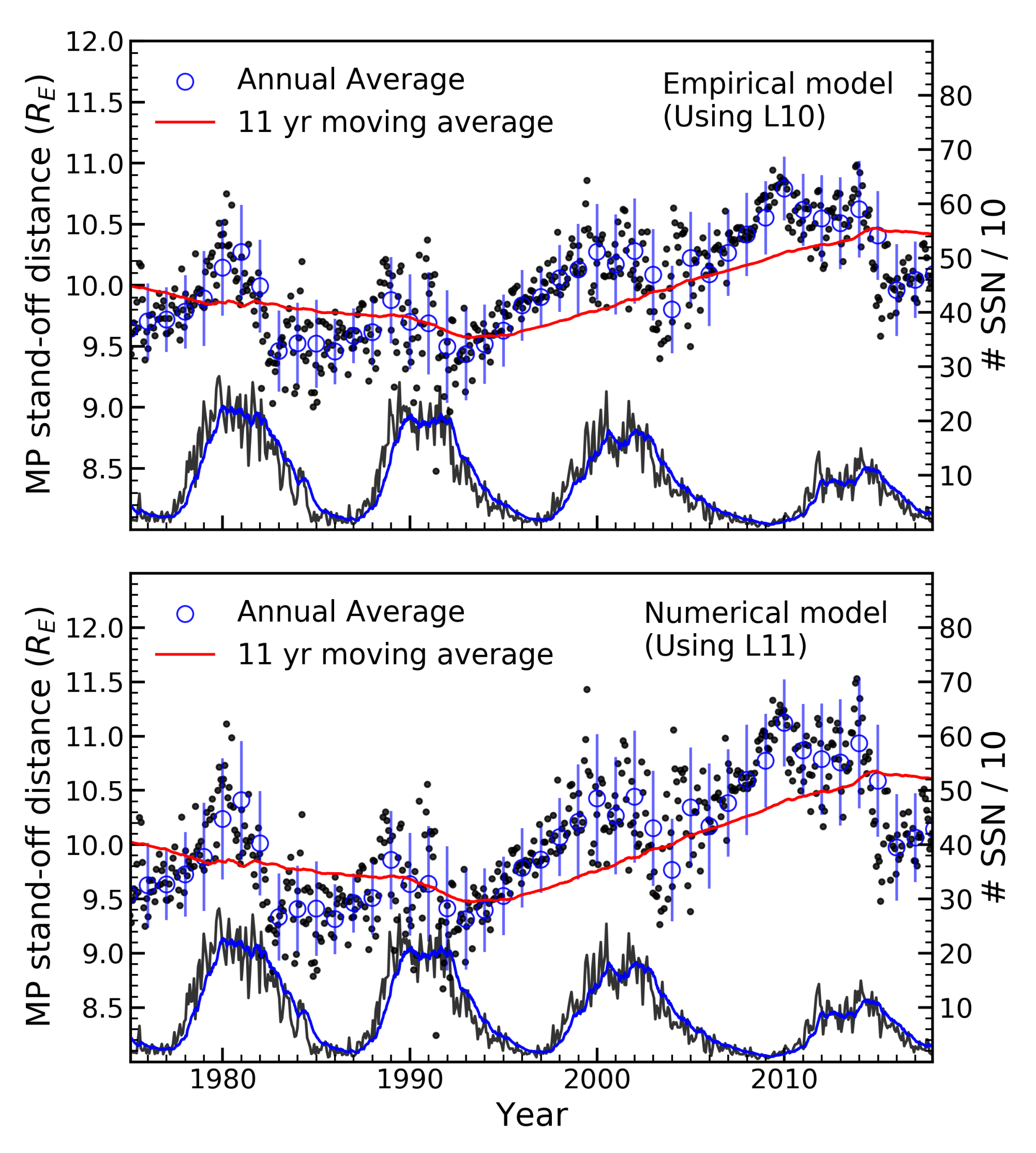}
		\caption{Carrington-rotation averages of MP stand-off distance
			for the period Feb. 1975$-$Dec. 2017  (grey filled
			circles) derived using \cite{LiZ10} (L10) (upper panel)
			and \cite{LuL11} (L11) (lower panel).
			The blue circles represent annual averages shown with
			1$\sigma$ error bars. The red line is a eleven year
			moving average of the daily average of MP stand-off
			distance. The monthly averaged sunspot numbers,
			scaled down by a factor of 10, is shown by the solid curve
			in grey with the smoothed values (one year moving
			average) overplotted in blue.}
		\label{fig6}
	\end{figure}
\end{center}
%

\section{Results} \label{sec:results}
It is to remind here that we discuss in the following
subsections about the temporal variations of daily averaged values
of MP and BS stand-off distances which directly obtained using
daily averaged IMF and solar wind plasma parameters in numerical
and empirical models as described in section 3.1 and section 3.2.
We further averaged the values of MP stand-off distances covering
one Carrington-rotation (27.2753 days) and also obtained the annual
averages to show the temporal variations in MP stand-off distance
and MP shape.
\subsection{Position of MP and BS}

\begin{figure*}[ht]
	\centering
	\includegraphics[width=0.55\textheight]{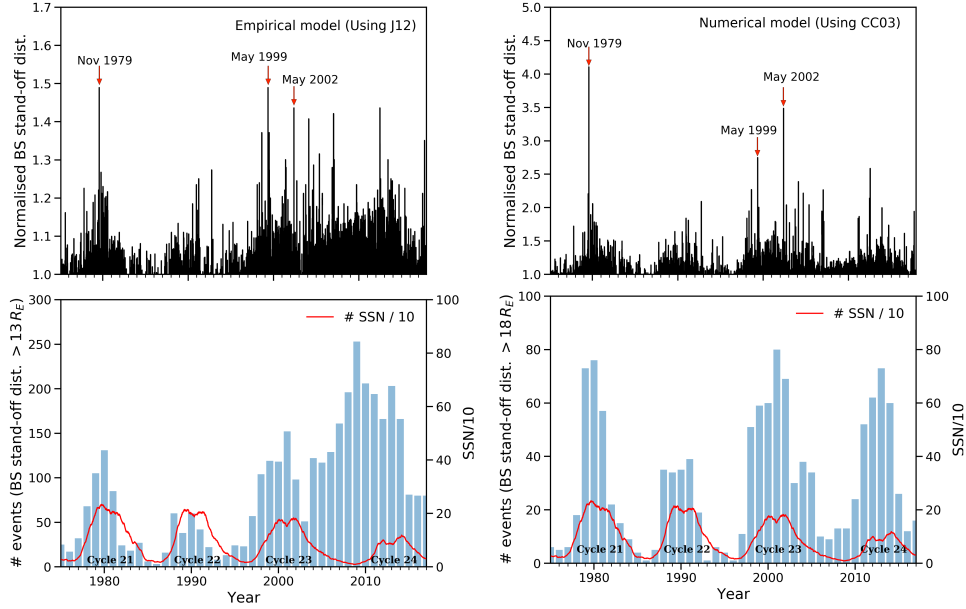}
	\caption{(Top) Daily averages of normalized BS stand-off
		distance between Feb. 1975 and Dec. 2017. Three
		extreme events, designated as solar wind disappearance
		events in the text, have been labelled with the event
		dates. (Bottom) The distribution of the number of
		events or instances for which the BS stand-off
		distance $>$ average stand-off distance between 1975
		and 2017. A 12 month moving average of sunspot numbers,
		scaled down by a factor of 10 is overplotted in red.
		Left panels uses J12 model while right panels uses
		CC03 model.}
	\label{fig7}
\end{figure*}

Figure \ref{fig6} shows the computed stand-off distance of the
MP during the period Jan. 1975$-$Dec. 2017. The top panel shows
the result obtained by using the empirical model L10 while the
results in the lower panel are derived using the numerical model
L11. The black filled dots are the Carrington-rotation averages
of MP stand-off distances, while the open blue circles represent
the annual means of MP stand-off distance shown with 1$\sigma$
error bars.  Shown at the bottom of each panel, by
a grey solid line overplotted in blue with the
smoothed value (of one year moving averages), is the monthly
averaged smoothed sunspot number, scaled down by a factor of 10.
It is clearly seen from both panels of Fig.\ref{fig6} that the
MP stand-off distance is modulated by the solar cycle activity,
having roughly a periodicity of 11 years. The comparison of MP
stand-off distance of each solar cycle reveals that the MP
stand-off distance starts to increase when the sunspot cycle is
roughly in the declining phase and it continues to increase until
the approach of the sunspot cycle maximum. This is much clearer
during the minima of solar cycles 22 and 23 when we see a clear
rise in stand-off distance. Besides this solar cycle modulation,
we can see a long term increasing trend in the MP stand-off
distances since the declining phase of solar cycle 22, around
the mid-1990's, in sync with the declining trend of the diminishing
sunspot cycles 22-24. To remove the 11-year solar cycle modulation
and investigate the increasing trend of MP stand-off distance,
the daily average of MP stand-off distance was smoothed using a
eleven year running mean, shown by the overplotted solid red
curve in both panels of Fig. \ref{fig6}. A clear increasing trend
in the MP stand-off distance is evident from the overplotted red
curve which begins to rise around the mid-1990's. This increase
was found to be $\sim15\%$, irrespective of the empirical or
numerical model used.
Figure \ref{fig7} shows the BS stand-off distances obtained by
using the empirical model (J12, \S{3.2.1}) (left panels) and
the numerical model (CC03, \S{3.2.2}) (right panels). The upper
panel of Fig.\ref{fig7} shows the daily averaged value of the
normalized BS stand-off distances from Feb. 1975 to Dec. 2017.
For Figure 7, we computed the daily BS stand-off distance after
normalizing by the average value of BS stand-off distance at
typical solar wind conditions at 1 AU (P${_{d}}$ $\sim$1.87 nPa,
B = 7 nT, N${_{p}}$ = 6.6 \textrm{cm}${^{-3}}$, ${\mathbf V \sim
440}$ km/s, M${_{a}}$ = 8). Thus, the normalized BS stand-off
distance can be less than unity. However, in Figure 7 (upper
panels), we presented only the normalized BS stand off distance
larger than unity. In this way, it is convenient to identify
the effects of the long term solar activity on the BS
stand-off distance, which is the main goal in this study. The
difference in the magnitude of the BS stand-off distance estimated
by the two models can be understood in the following way. Both
the models, CC03 and J12 relate the BS stand-off distance with
solar wind dynamic pressure as a power law, but with different
power law indices. Also, the CC03 model includes the effect of
the Alfv\'en Mach number (M${_{A}}$), which has been neglected
in J12.
%
%
%
%
It is clear that irrespective of the models used, the normalized BS
stand-off distances show a large number of cases of the BS stand-off distance extending well beyond the average value.  Three such events are indicated in Fig.\ref{fig7} (upper panel). These events have been well studied and are referred to as solar wind disappearance events \citep{BaJ03,JaF05,JaF08,JDM08} due to the extremely low densities observed at 1 AU ($<$ 0.1 $\textrm{cm}^{-3}$) for periods exceeding 24 hours. During all three solar wind disappearance events, a sharp decrease in $P_d (< 0.02 \textrm{nPa}$) was seen indicating the sensitive response of the BS stand-off distances to the changed solar wind conditions. Also, it is seen from Fig.\ref{fig7} (upper panel) that the normalised BS stand-off distance, on an average, follows the 11-year solar cycle. However, a careful inspection of the
model results shows that the solar cycle modulation is very clear
when we used the model by CC03. While the results obtained by using
J12 show large excursions of the normalised BS stand-off
distance extending well below the average value during cycle 23
minimum. It is to be noted that solar cycle 23 witnessed an unusual and deep solar minimum during the same period.
It appears that there is a difference in the model results for
the normalized BS stand-off distance during cycle 23 with the
empirical model calculations showing more number of days having
an increased BS stand-off distance than the numerical model
calculations.

To further quantify the effect of solar wind conditions on the
BS, we selected those days with normalized values of BS stand-off
distance which are more than $1 \sigma$ above the average BS stand-off
distance at typical solar wind conditions, which we referred to as an
event. For typical solar wind conditions at 1 AU, the average BS
stand-off distance according to J12 is $13 R_{E}$ and that according
to CC03 is $18 R_E$.
The lower panels of Fig.\ref{fig7} shows the histogram of the distribution of the number of events for the period 1975--2017 obtained using the empirical model by J12 (lower left) and the numerical model by CC03 (lower right). As stated earlier, it is clear that the BS excursions beyond the average BS stand-off distance are observed consistently for each solar cycle and we see a clear modulation of solar cycle for solar cycles 21--23, irrespective of the model used. In order to show the solar cycle modulation, we overplotted, in red, the 12-month moving average of the smoothed sunspot number, scaled down by a factor of 10, for the period 1975--2017, as shown at the bottom of each of the lower panels of Fig.\ref{fig7}.  As pointed out earlier, we see a difference
in the number of events during the solar cycle 23 minimum for the
results obtained by the empirical model J12.  Also, evident from
the lower panels of Fig.\ref{fig7} is that there is a
significant increase in the number of events with the BS
stand-off distance exceeding well below the average value since
around mid-1990's. One needs to bear in mind, that at
the same time we have seen solar photospheric magnetic fields
showing a steady decline. In case of both the models, the
increase, that we see in the number of events after mid-1990's
is found to be more than $\sim$ 40\% when compared with the
number of events before mid-1990's.
\subsection{Shape of MP}
\begin{figure}[ht]
	\centering
	\includegraphics[width=0.4\textheight]{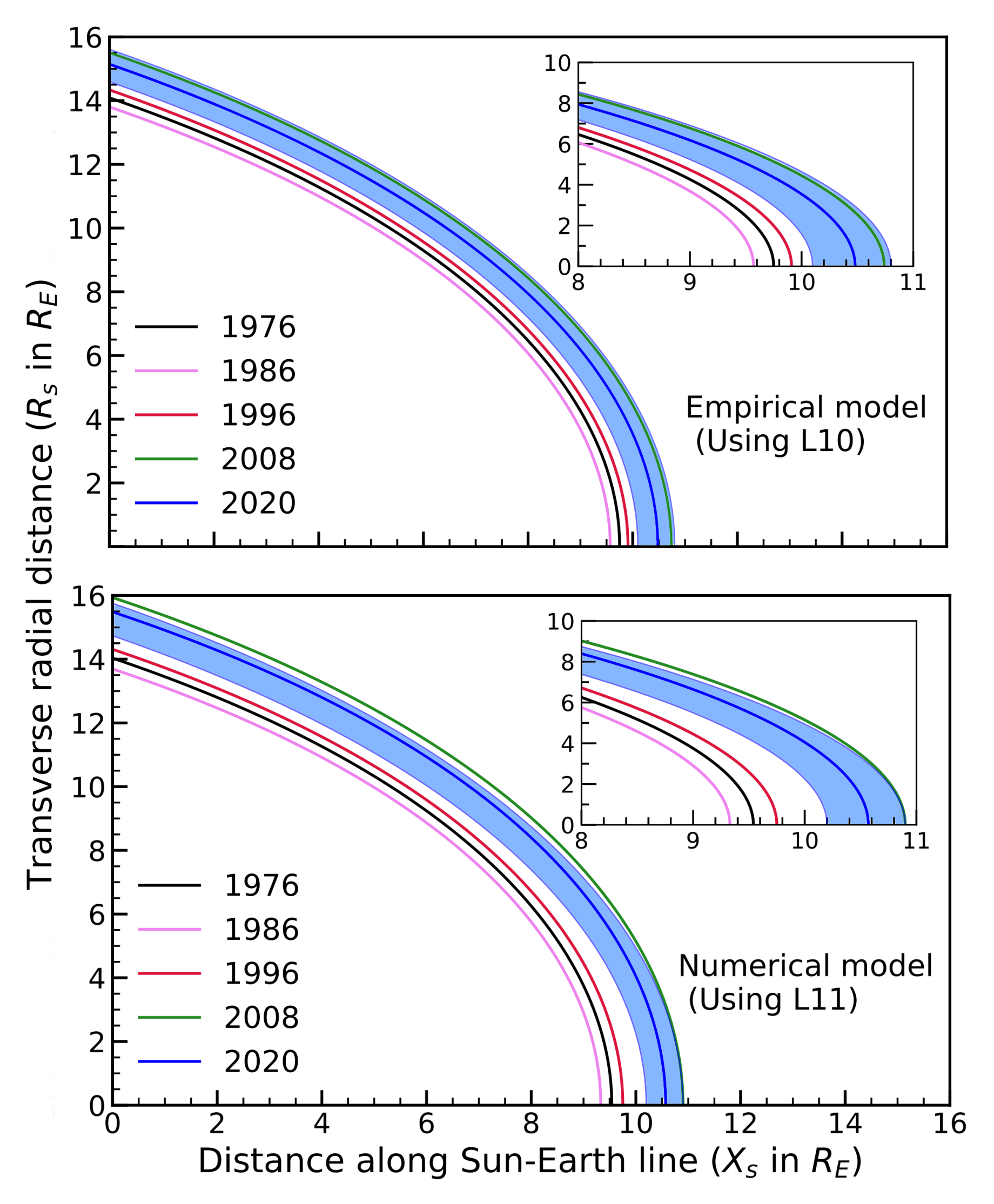}
	\caption{The one year averaged MP shape at solar minima
		for cycles 20-24 shown by a plot of transverse radial
		distance of MP, $R_s$ against the extent of the MP
		along the sun-earth line, $X_s$. The upper panel
		uses \cite{LiZ10} (L10) (\S{3.1.1}) and the lower
		panel uses \cite{LuL11} (L11) (\S{3.1.2}). The solid
		black line is the one year interval averaged of MP
		shape for the year 1976 during the minimum of cycle 20. Similarly, the pink, red and green solid curves represent
		1 year averaged MP shapes for years 1986, 1996 and
		2008 during solar minima of cycles 21, 22 and 23,
		respectively. The blue curve presents the forecast of
		MP shape in the year 2020, with a shaded blue band of
		95\% confidence interval, for the minimum of the current
		cycle 24. Also, an inset plot presenting the zoomed in
		MP shape is shown in both upper and lower panels.
	}
	\label{fig8}
\end{figure}

Using the standard GSM coordinate system we computed the position of
MP, $X_s = r cos(\theta)$ (the stand-off distance along the
Earth-Sun line for the day-side MP) and $R_s = r sin(\theta)
= \sqrt(Y^2 + Z^2)$ (in GSM coordinate, also known as the
transverse radius), where $\theta$ is the solar zenith angle.
In order to compute the shape of MP at solar minima for cycles
20--23, we plotted $R_s$ v/s $X_s$, averaged over 1 year intervals
for years 1976, 1986, 1996 and 2008 during each solar minimum for
cycles 20, 21, 22 and 23, respectively. The 1-year interval
corresponds to the period shown by grey vertical bands in Fig.\ref{fig2}. We refer to the plot of $R_s$ v/s $X_s$ as the
MP shape. Figure \ref{fig8} shows the plot of $R_s$ v/s $X_s$ for
a solar zenith angle between $\theta$= 0${^{\circ}}$ and
$\theta$ = 90${^{\circ}}$. The upper panel of Fig.\ref{fig8}
shows the MP shape obtained using the eq. (\ref{mp_shape_L10})
(\S{3.1.1}) based on the empirical model by L10, while the lower
panel shows the MP shape obtained using the eq. (\ref{mp_shape_L11})
(\S{3.1.2}) based on the numerical model by L11. The 1-year
averaged MP shape is labelled by the year, for {\it{e.g.}} the
MP shape of 1976, shown by the solid black curve, refers to the
MP shape that is averaged over 1 year in 1976 during the solar
minimum of cycle 20. Similarly, the solid curves in indigo, red
and green represent 1 year averaged MP shape for years 1986, 1996
and 2008,  for solar minima of cycles 21, 22 and 23, respectively.
Also, an inset plot presenting the zoomed in MP shape is shown
in each panel of Fig.\ref{fig8} that shows clearly the expansion
of MP shape. It is evident from Fig.\ref{fig8} that in both cases
(for L10 (upper panel) and L11 (lower panel) models) the average
MP shape in 1986 for the solar minimum of cycle 21 falls below the
average MP shape in 1976 for the solar minimum of cycle 20,
and then bounces forward in 1996 and continues to expand until
2008. So the 1 year averaged MP shape during solar minima of
cycles 22--23 showed a continuous expansion.

For the period between 1996 and 2008, the stand-off point
expanded by nearly 1 R${_{E}}$ from $\sim$9.8 R${_{E}}$ to
$\sim$11 R${_{E}}$, irrespective of the model used. We have
also computed the MP flaring angle that is a function of
$P_d$ and IMF-$B_z$ \citep{LuL11}. We have found that the
flaring angle also shows a 11 year solar cycle modulation
(Figure not shown here) and when this modulation is removed
by smoothing the flaring angle with a 11 year moving average,
we see a clear increasing trend since 1996.
%
%
%

Under the possibility of having a continuing declining trend
in solar polar fields at least until 2020, it is imperative
to know how the Earth's magnetosphere will look like in 2020.
So we forecast the shape of the MP for the year 2020 during the
forthcoming minimum of cycle 24. As stated earlier, the MP shape
is sensitive to the changes in P${_{d}}$ and $B_z$
(see equations \ref{mp_shape_L10} and \ref{mp_shape_L11}).
We, thus, carried out a systematic exercise
to identify the trend and seasonality in the time series data
of P${_{d}}$ and $B_z$ using a Auto Regressive Integrated
Moving Average (ARIMA) process (see Appendix A). This
enables us to obtain a stationary time series facilitating
further analysis and to make a forecast. The forecast is
obtained for monthly averages of $B_z, P_d$ and $r_{mp}$ for
the time period between Dec. 2017 and Dec. 2020 (see Figures
in Appendix A). We compared the predicted values of $P_d$
and $r_{mp}$ by the ARIMA method with the observed values of
$P_d$ and the estimated values of $r_{mp}$ obtained from the
OMNI2 data base between the period 2018-01-01 and 2019-03-01.
It is evident from the forecast of $P_d$ and $r_{mp}$
that the predicted and the observed values between the period
2018-01-01 and 2019-03-01 are in good agreement. This
verifies the ARIMA method. It is, however, not the case for
$B_z$ which shows a simple linear extrapolation
curve without any variations. However, using the ARIMA method
the value of $B$ in 2020 has been estimated to be $\sim$5.4
($\pm$0.9) nT. Using the correlation between B and solar polar
fields, we also estimated a value of 5.3 ($\pm$0.3) nT at 2020
for the value of $B$. The values of B computed from these two
separate methods agree well within uncertainty.
This further validates the forecast using the ARIMA method.
The curve in blue in Fig.\ref{fig8} depicts the forecast of the
MP shape for 2020 for the minimum of cycle 24. The blue band
in Fig.\ref{fig8} signifies the region of 95\% confidence
interval around the MP shape of 2020. It is clearly seen
from Fig.\ref{fig8} (both upper and lower panels) that the
MP shape in 2020 shows further expansion than in 1996 (during
solar minima of cycle 22). However, it resumes an average
value of 10.3 $R_E$, which is less than that in 2008 (during
solar minima of cycle 23).
\section{Discussion}
In the present study, we quantified the response of the earth's
magnetosphere via the variations in the stand-off distance of the
MP and BS for the period February 1975--December 2017 caused by the
long term changes that have been seen in solar photospheric
magnetic fields, inner heliospheric solar wind micro-turbulence
levels and solar wind changes at 1 AU, since the mid-1990's.
We carried out the computation of MP and BS stand-off distances
as well as the MP shape using both empirical and numerical models
and compared the model results.  The MP stand-off distance and
the MP shape was computed using L10 (empirical model, \S{3.1.1})
and L11 (global MHD simulations, \S{3.1.2}). For the BS
stand-off distance we used the model by CC03 (based on
global MHD simulations, \S{3.2.2}) and compared them with the
results obtained by J12 (empirical model, \S{3.2.1}).

Our analysis showed a continuation of the steady decline until 2017
in the high-latitude photospheric magnetic fields and solar wind
micro-turbulence levels in the inner heliosphere. Also, a similar
behaviour of global reduction in values of solar wind parameters
at 1 AU such as B, N${_{p}}$, v, P${_{d}}$ and He$^{++}$/H$^{+}$
post mid-1990's was clearly seen. Corresponding to this observed
decline, a steady increase of $\sim$15\% was observed in the
stand-off distance of the MP, irrespective of the model used. Our
result of the increase in the MP stand-off distance is consistent
with the increase in the stand-off distance of the MP reported
by \cite{McA13} wherein, the authors found an increase in the MP
stand-off distance from 10 $R_E$ in 1974$-$1994 to 11 $R_E$ in
2009--2013. From our study, the value of the stand-off distance of
MP from 1974$-$1994 were found to be 9.7 R${_{E}}$. In contrast, the
value of MP stand-off distance from 1995$-$2017 were found to be
10.7 R${_{E}}$. So \cite{McA13} and our study show a similar increase
of about 1 $R_E$ in MP stand-off distance. We also observed a
significant increase of more than 40\% in the number of events
(after 1995) where the stand-off distance of the BS exceeded the
average stand-off distance over the past $\sim$24 years, when
compared with the number of events prior to 1995.

It is to be noted that both P${_{d}}$ and IMF magnitude
have shown a similar temporal behaviour \citep{RWP01, McE03,
JRL11, McA13} with a global reduction of $\sim$40\% in their
average values over the past 24 years. In our analysis, we
showed this behaviour of P${_{d}}$ and IMF that is continuing
till date. Generally, during solar minimum, photospheric high
latitude magnetic fields extend to low latitudes and are then
pulled into the heliosphere by the solar wind thereby, forming
the IMF \citep{SPe93}. The changes in the solar wind conditions
such as decline in P${_{d}}$ and IMF strength can thus be
interpreted as being induced by global changes in solar
photospheric magnetic fields. We also showed this causal link
of P${_{d}}$ and IMF with solar polar fields in our analysis. The
computed correlations of P${_{d}}$ and IMF with solar polar fields,
B${_{p}}$ over the past four solar cycles apparently proved this
connection between them, which has been found to be better if
the correlation is considered for only the minima of cycles 23
and 24, {\it{i.e.}}, post mid-1990's (with correlation
coefficients of 0.95 and 0.92 for P${_{d}}$ and IMF with B${_{p}}$, respectively).
A decrease in P${_{d}}$ and IMF causes an expansion of the BS and
MP, resulting in an increase in their sub-solar distances. The
stand-off distance of the MP, in general, exhibits a power law
dependence on the dynamic pressure, with power law index $\sim-1/6$
\citep{MBe64, PRu96}. A self-similar scaling suggests an identical
power law dependence for the stand-off distance of the BS \citep{CLy96}.
As mentioned earlier, the models used for the computation of the
stand-off distances of BS and MP are parametrized by the changes in
both P${_{d}}$ and the IMF, so a global rise in the stand-off distance
of MP and BS is thus expected considering the global reduction in both
P${_{d}}$ and the IMF.  We checked the values of P${_{d}}$ from
1974$-$1994 and 1995$-$2017, which were found to be, respectively,
$\sim$2.9 nPa and $\sim$2.0 nPa, showing a decrease of $\sim$~31\%
post-1995 in comparison to pre-1995. The corresponding increase of
$\sim$9.3\% observed in the stand-off distance of the MP can thus
be attributed to the power law dependence of the MP stand-off distance
on the P${_{d}}$. However, the power law index found in several
empirical as well as numerical studies is little less than $-1/6$
\citep{LiZ10, ShS98}, which is the theoretically
predicted value for a dipole in vacuum. This deviation from
$-1/6$ can probably accounted by a plasma pressure inside the
magnetosphere \citep{JNS12}.
The increasing trend in stand-off distances of the MP and BS is
therefore consistent with the ongoing declining trend in high-
latitude photospheric magnetic fields and solar wind micro-turbulence
levels that we have seen during this last 22 years. This indicates
that the earth's magnetosphere is very sensitive to the changes in
solar wind conditions modulated by changes in solar magnetic activity.

From our study, we found that the strength of IMF and P${_{d}}$
averaged over 1 year intervals during solar cycle minima showed
a better correlation with solar polar fields.  Based on this
correlation, we also studied the variation in the average
MP shape for several 1-year intervals during solar minima of
cycles 20--23 and found that the average MP shape in 1996 during
the minimum of cycle 22 has shown an outward expansion, which
has undergone a further significant expansion in 2008 during
the minima of cycle 23, which happened to be one of the deepest
experienced in the past 100 years. Our results showed an average
increase in the MP shape of $>$ 1 $R_E$ for the period 1996$-$2008.
Based on the assumption that the steady decline in solar high-latitude
fields would continue, we predicted the shape of MP for the
expected minimum of cycle 24 in 2020 using the forecast values
of $B_z, P_d$ and $r_{mp}$ for the time period between Dec. 2017
and Dec. 2020 obtained by the ARIMA method. The predicted shape
of the MP in 2020, during the expected minimum of solar cycle 24,
suggests an expanded terrestrial magnetosphere with
the MP shape in 2020 being larger than the MP shape during the
minimum of cycle 22 and smaller than the MP shape during the
minimum of cycle 23.

The comparison of the model results for the long term trend in the
MP stand-off distance and shape, in general, show they do not depend
on the model used. While the comparison of the model results for
the long term trend in the BS stand-off distance also show that
they are independent of the model used for solar cycles 21--22.
However, they show a difference for the BS stand-off distance
for the period of solar cycles 23--24. 

The location of the MP inside the position of geostationary
orbit is crucial since the MP acts as a shield
against high speed solar wind during geomagnetic
storms caused by CME impacts on the magnetosphere. These impacts,
if they originate from a large high velocity CME have the ability
of compressing the MP to distances below the geostationary orbit,
which lies at a distance of around $6.6 R_E$, thereby exposing
the geo-stationary orbit to high energy particle flux from the
sun and posing a threat of causing damage to or destroying
satellite systems entirely. We list in Table \ref{tab-MP} the
events when the daily values of MP stand-off distance
comes close to the geostationary orbit ($\sim 6.6 R_{E}$),
computed using the L10 and L11 models separately. The solar wind
dynamic pressure corresponding to the MP stand off distance = 6.6 R${_{E}}$ is approximately 23 nPa. Form Table \ref{tab-MP}, it
is seen that there were two events between the year
1968 and 1995 in contrast to the single event post 1995.
The data also showed four other events between 1968
and 1995, but since there were large data gaps in these data sets,
they have not been included in Table \ref{tab-MP}.
The less number of events with MP stand-off distance
less than $7.0$ R${_E}$ post mid-1990's agrees with the observed
increasing trend of MP stand-off distances post mid-1990's.
\begin{table}
	\begin{center}
		\caption{Daily averaged values of MP Stand-off
				distance ($r_{mp}$ in $R_E$) closest to
				geostationary orbit ($6.6$ R${_E}$) using models
				due to \cite{LiZ10} (L10) and \cite{LuL11} (L11)}
		\label{tab-MP}
		\vspace{0.1cm}
		\begin{tabular}{lccc}
			\hline\noalign{\smallskip}
			& $r_{mp}$ in $R_E$ & $r_{mp}$ in $R_E$ & Duration \\
			Date & Using L10		 & Using L11 & ($\sim$hours)  \\
			\noalign{\smallskip}\hline\noalign{\smallskip}
			1978-06-02 & 6.62 $\pm$ 0.16 & 6.31 $\pm$ 0.18 & 11 \\
			1991-06-05 & 6.83 $\pm$ 0.20 & 6.51 $\pm$ 0.22 & 10 \\
			2005-01-17 & 6.99 $\pm$ 0.15 & 6.84 $\pm$ 0.16 & 09 \\
			\noalign{\smallskip}\hline
		\end{tabular}
	\end{center}
\end{table}

Our study shows the steady decline in high-latitude photospheric fields and solar wind micro-turbulence levels have been continuing
until December 2017, the end of our data set. On the assumption that
the decline would continue until 2020, the expected minimum of the
current solar cycle 24. \cite{JaB15} predicted that the next solar
cycle 25 would be a weaker cycle with a maximum smoothed sunspot number (SSN) of 62$\pm$12 on the old unrevised scale of sunspot number
(V1.0). Other researchers have also made similar predictions of
a weaker solar cycle 25 \citep{CJS16,UHa18,BNa18}.
Based on the hemispheric asymmetry of solar polar field reversal
in cycle 24 and the prolonged, 2.5 year long, zero-field condition
before the completion of polar field reversal in the solar
northern hemisphere, \cite{JaF18} predicted a weaker cycle 25, similar
to cycle 24.  In a paper (Bisoi et al., submitted manuscript, 2019),
also predicted a weaker cycle 25 with a maximum SSN of 132$\pm$11 on
the revised scale of sunspot number (V2.0), which is slightly stronger
than cycle 24. Based on the continuous declining trends of solar
high-latitude fields for the past 25 years and the prediction of a
weaker cycle 25, we expect a low level of sunspot activity in future
beyond solar cycle 25, if the declining trends continue beyond 2020, a
condition akin to the Maunder minimum (1645-1715) when the sunspot
activity was extremely low or almost non-existent. Other researchers
have also arrived at similar conclusion and reported that the Sun may
move into a period of very low sunspot activity comparable with the
Dalton minimum \cite{ZPo14}, while \citep{ZGk15,San16} claim that
the Sun is moving into another extended Maunder-like minimum. However,
we have not carried out any extensive analysis to show the long term forecasts of Maunder minimum like conditions. Our interpretation is purely based on the assumption of a continuation of declining trend of solar high-latitude photospheric fields. The current understanding
is that sunspot fields are generated by a shearing of pre-existing poloidal fields (high latitude fields) and the regeneration of poloidal fields through a Babcock-Leighton mechanism with the whole process operated by an interior solar dynamo. The strength of poloidal fields depends on the large scale scatter in the sunspot fields and the tilt angle distribution of bipolar sunspot regions \citep{DCh93}. This is a random process. We have seen continuing weaker polar fields in cycles 22 and 23. The same is also seen in cycle 24 that is continuing till date. The probability for a random process of polar field regeneration yielding three successive weaker cycles is clearly
very low and therefore not only extremely unusual, but points to a
lack of a complete understanding of the solar dynamo process. Thus,
we expect a continuation of the decline in solar high-latitude fields
in future solar cycles, that could result in very low sunspot
activity in future and possibly result in conditions akin to the
Maunder minimum. However, the long term decline in solar fields
could represent a period of solar activity with a period longer
than 11 years.  As of now, we have solar magnetic field
measurements only for the last four solar cycles,
cycles 21--24, it is difficult to determine
the longer periods than the known period of 22 years (or a solar
magnetic cycle) in solar high-latitude magnetic fields.  Also,
the solar high-latitude fields have been showing a long term decline
more than 25 years (greater than the solar magnetic cycle of 22
years). We, therefore, expect a continuation of the decline
of high-latitude fields beyond solar cycle 24.

Using a global thermodynamic model, \cite{RiL15}
reported the likely state of the solar corona, during the later
period of the Maunder minimum, devoid of any large scale structure
and driven by a reduced photospheric magnetic field strength.
The photospheric field strength during the last two solar cycles
has been steadily decreasing (since $\sim$1995) and the trends
indicate that it is likely to decline in the same manner in
future solar cycles. This implies a state of corona with no large
scale structure much alike Maunder minimum period, which in turn,
leads to an expansion of terrestrial magnetosphere
with an increased stand-off distance for the BS and the MP.

%


%
%
%
%
\section{Conclusions}

Our study underlines the causal relationship between
solar activity changes and the corresponding global
response of the earth's magnetosphere via the variations
quantified by the stand-off distances of the BS and MP.
Our study mainly showed an increasing trend in the
stand-off distance of MP and BS corresponding to the
observed decreasing trend in solar high-latitude photospheric
polar fields and solar wind micro-turbulence levels since the
mid-1990's. A similar increasing trend was also observed
for the MP shape till 2008 which is the minimum of the
solar cycle 23. However, a forecast of the MP shape in
2020, the expected minimum of cycle 24, showed a smaller
MP stand-off distance. Further, we find two instances
between 1968 and 1991 when the MP stand-off distance
reach closest to 6.6 Earth radii (the geostationary orbit)
for duration ranging from 9$-$11 hours
and a single event in 2005, after the start of the decline
in photospheric fields began. Finally, the decline in
photospheric fields has now continued for over two solar
cycles or one full magnetic cycle of 22 years. If the
declining trend continues beyond 2020, a Maunder minimum
like condition may be expected beyond solar cycle 25.

Continued investigation to understand and forecast the
influence of solar activity on the near earth environment
and the ecosystem is, therefore, of considerable importance.

\appendix
\section{: Forecasting a time series}
Auto-Regressive Integrated Moving Average, ARIMA$(p, d, q)$,
is a general class of models for forecasting a stationary time
series. A non-seasonal ARIMA forecasting equation for a time
series is a linear equation given by \citep{Bro17},

 	 	\begin{equation}
 	 	{Y_t} = {\mu} + Y_{t-d} + {\sum}_{i=1}^{p} {\phi}_i Y_{t-i}-
 	 	{\sum}_{j=1}^{q} {\theta}_j {\epsilon}_{t-j}
 	 	\end{equation}

Where, $\mu$ is a constant, second term represents $d$ number
of non-seasonal differences needed to make time series stationary.
Third term is a weighted $(\phi_i)$ auto-regressive process
of the order of $p$ and fourth term corresponds to a weighted
$(\theta_i)$ average of the noise (moving average process) of
the order of $q$.

	To forecast the shape of the MP based on the models
	due to L10 and L11, we need to forecast two parameters, namely
	- solar wind magnetic field $(B_z)$ and dynamic pressure $(P_d)$.
	Using the forecast for $B_z$ and $P_d$, we can
	derive the forecast of the MP stand-off distance $(r_{mp})$. This
	then yields the shape of the MP \\

	\noindent Following steps are involved in obtaining a forecast
	\citep{Bro17}:
	\begin{enumerate}
		\item Decomposition of time series
		\item Analysis of stationarity
		\begin{enumerate}
			\item Moving average and standard deviation
			\item Augmented Dicky-Fullar test
		\end{enumerate}
		\item Making a time series stationary
		\item Determine the parameters of ARIMA
		\begin{enumerate}
			\item Auto-correlation function (ACF) and Partial
			Auto-correlation function (PACF)
			\item Akaike information criterion (AIC)
		\end{enumerate}
		\item Time series forecast using the best fit parameters
		determined in step 4.
	\end{enumerate}

	The decomposition of the time series ($B_z$
	and $P_d$) into trend, seasonal variations and residue has
	been carried out. We found that $P_d$ and $B_z$ shows
	modulation in response to 11 year sunspot cycle. 11 year
	moving average reveals an increasing trend in $P_d$, with
	seasonal variations of the period of one year. $B_z$, on the
	other hand, shows dominant stochastic behaviour throughout.
	We applied Seasonal ARIMA (SARIMA) method to $B_z$ and $P_d$
	in order to obtain the forecast. This requires to estimate
	the hyper-parameters P, D and Q corresponding to the seasonal
	components of the parameters of ARIMA (p, d, q). This is
	generally represented as $(p, d, q) \times (P, D, Q, m)$
	where m is the period of seasonal variations.

	The analysis of stationarity shows that all three parameters
	$(B_z, P_d$ and $r_{mp})$ follow 11 year sunspot cycle. Also
	11 year moving average shows decreasing trend in $B_z$ and
	$P_d$ and increasing trend in $r_{mp}$ since $\sim 1995$. We
	considered Augmented Dicky-Fullar test (ADF) to determine the
	stationarity of the time series. In all three cases we find
	that after applying first order difference we get test
	statistics (ADF-test) well below the 5\% critical value with
	no trend in 11 year moving average and standard deviation of
	the time series. Thus the first order difference time series
	of $(B_z, P_d$ and $r_{mp})$ is considered as stationary to
	carry out further analysis and obtain the forecast.

\begin{figure}[ht]
	\centering
	\includegraphics[width=0.4\textheight]{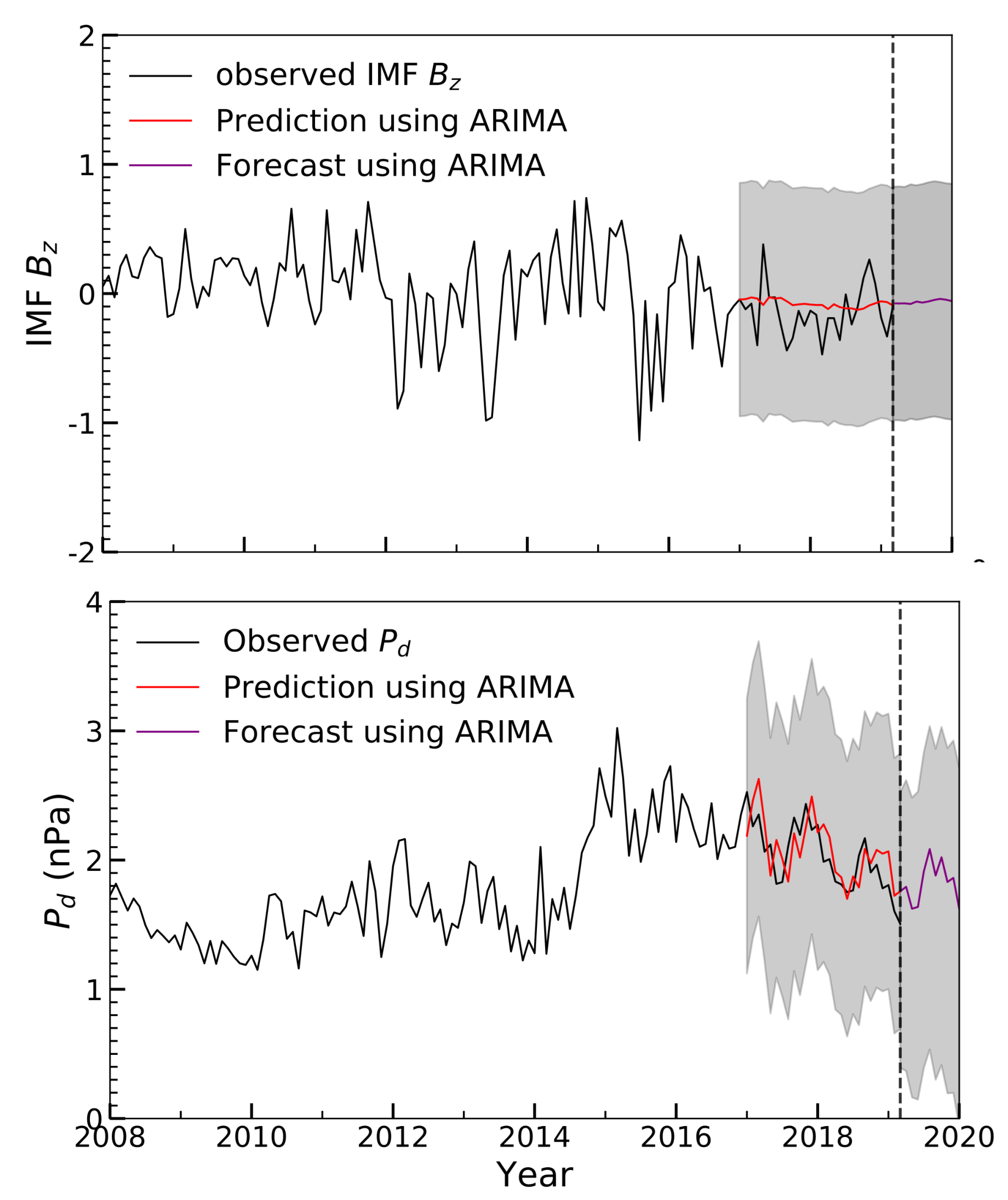}
	\caption{The forecast of IMF-$B_z$ (upper panel) and $P_d$ (lower
	panel). The solid line in black represents the actual monthly
	averaged values of IMF-$B_z$ and $P_d$, while the solid red and
	violet lines are the predicted and forecast values of IMF-$B_z$
	and $P_d$ obtained using the ARIMA method for the
	period between 2018-01-01 and 2019-03-01 and between 2019-03-02
	to 2020-12-31, respectively. The grey band shown is the 95\% confidence interval. The vertical line demarcate the
	time period from when we have shown the forecast values.}
	\label{fig9}
\end{figure}

	Next we need to determine the ARIMA parameters $p, d$ and $q$
	and their seasonal components P, D and Q as defined above. We
	followed ACF and PACF analysis to get the range of most
	likely parameters and carry out grid search. Based on the AIC
	criterion we then determined the best fit parameters that
	turns out to be $(1, 0, 1)\times(0, 0, 1, 12)$ for $B_z$ and
	$(1, 1, 1) \times (0, 0, 1, 12)$ for $P_d$ corresponding to
	the lowest value of AIC. Using these parameters the forecast
	for the time series of $B_z$ and $P_d$ is obtained as shown
	in Fig.\ref{fig9}. The solid black line represents the actual
	monthly averaged values of IMF-$B_z$ and $P_d$ obtained from
	the OMNI2 data base. The solid red and violet lines are
	forecast values of IMF-$B_z$ and $P_d$ obtained using the
	ARIMA method. We compared the predicted values of $P_d$ and IMF-$B_z$ by ARIMA method with the observed values of $P_d$
	and IMF-$B_z$ obtained from OMNI2 data base between the period 2018-01-01 and 2019-03-01. It is clearly evident from the
	forecast of $P_d$ that the predicted and the observed values between the period 2018-01-01 and 2019-03-01 are well in agreement. This is, however, not the case of $B_z$ which
	rather shows a simple linear extrapolation curve without any variations indicating that the IMF-$B_z$ does indeed stochastic
	in nature.
%
%

\begin{figure}[ht]
	\centering
	\includegraphics[width=0.4\textheight]{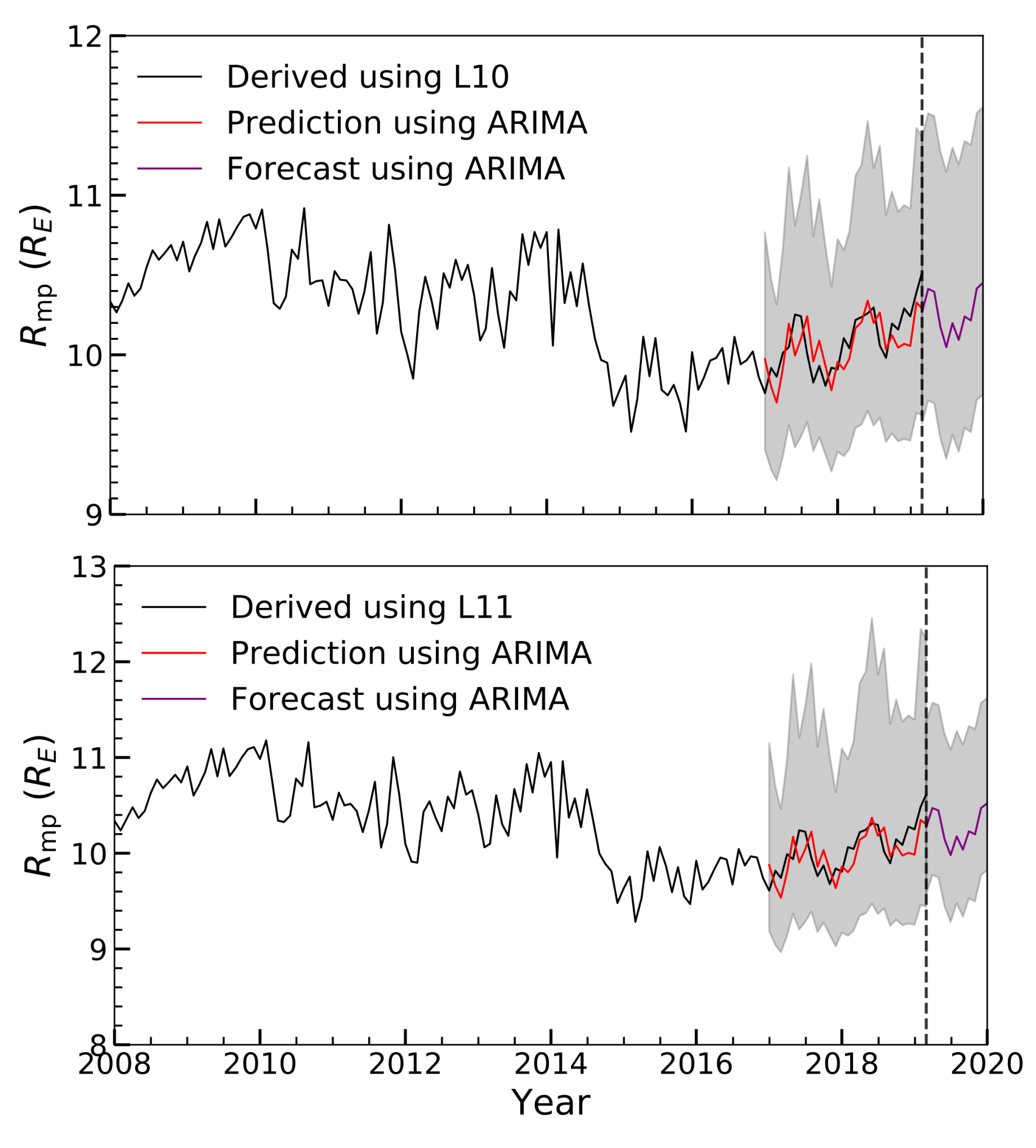}
	\caption{The forecast of $r_mp$. The upper panel uses
	\cite{LiZ10} (L10) (\S{3.1.1}) and the lower panel uses \cite{LuL11} (L11) (\S{3.1.2}) models. The solid line in
	black represents the actual monthly averaged values of
	$r_mp$, while the solid red and violet lines are the
	predicted and forecast values of $r_mp$ obtained using
	the ARIMA method for the period between 2018-01-01 and
	2019-03-01 and between 2019-03-02 to 2020-12-31,
	respectively. The grey band shown is the 95\% confidence
	interval. The vertical line demarcate the time period
	from when we have shown the forecast values.}
	\label{fig10}
\end{figure}

	The forecast of $B_z$ and $P_d$ is then corrected for the first order difference and implemented in the models (L10 and L11) to compute the forecast of $r_{\mathrm{mp}}$ as shown in Fig.\ref{fig10}. Thereafter, the MP shape for 2020 is obtained
	as shown in Fig.\ref{fig8}.  Fig.\ref{fig10} shows the forecast
	of $r_{\mathrm{mp}}$ using the ARIMA method. The solid black
	line is the calculated $r_{\mathrm{mp}}$ directly using the observed IMF-$B_z$ and $P_d$ from OMNI2 data base. The solid
	red and violet lines are predicted values of $r_{\mathrm{mp}}$.
	We compared the predicted values between 2018-01-01 and
	2019-03-01 with the estimated values of $r_{\mathrm{mp}}$
	directly from the IMF-$B_z$ and $P_d$. It is evident from Fig.\ref{fig10} that the predicted and the estimated values
	of $r_{\mathrm{mp}}$ between the period 2018-01-01 and
	2019-03-01 are well in agreement. For the prediction of $r_{\mathrm{mp}}$, we have used the predicted non-zero value
	of IMF-$B_z$, as shown Fig.\ref{fig9}. However, we have
	carried out the computation by assuming $B_z$ = 0 and found
	no significant change in $r_{\mathrm{mp}}$ compared with
	the computations using a finite non-zero value of IMF-$B_z$
	(for example, for $r_{\mathrm{mp}}$ at 2020, we used $B_z$=0.038).

%

From Fig.\ref{fig9} and Fig.\ref{fig10}, we see that the predicted
values of $P_d$ and $r_{\mathrm{mp}}$ by ARIMA method agree well
with that of the observed values obtained from OMNI2 data base.
This verifies the forecast of MP shape using the ARIMA method.

\acknowledgments
This work has made use of NASA's OMNIWEB services Data System.
The authors thank the free data use policy of the National Solar
Observatory (NSO/KP, NSO/SOLIS and NSO/GONG), OMNI2 from NASA
and WDC-SILSO at Royal Observatory, Belgium, Brussels. JP acknowledges
the ISEE International Collaborative Research Program for support
during this work. SKB acknowledges the support by the PIFI (Project
No. 2015PM066) program of the Chinese Academy of Sciences
and the NSFC (Grants No. 117550110422, 11433006, 11790301, and 11790305).


%
%




\end{document}